# Power Consumption of Virtualization Technologies: an Empirical Investigation

Roberto Morabito[1,2]
[1]*Ericsson Research, NomadicLab,* Jorvas, Finland
[2]*Department of Communications and Networking (Comnet), Aalto University,* Espoo, Finland
roberto.morabito@ericsson.com

*Abstract*— Virtualization is growing rapidly as a result of the increasing number of alternative solutions in this area, and of the wide range of application field. Until now, hypervisor-based virtualization has been the de facto solution to perform server virtualization. Recently, container-based virtualization – an alternative to hypervisors – has gained more attention because of lightweight characteristics, attracting cloud providers that have already made use of it to deliver their services. However, a gap in the existing research on containers exists in the area of power consumption. This paper presents the results of a performance comparison in terms of power consumption of four different virtualization technologies: KVM and Xen, which are based on hypervisor virtualization, Docker and LXC which are based on container virtualization. The aim of this empirical investigation, carried out by means of a testbed, is to understand how these technologies react to particular workloads. Our initial results show how, despite of the number of virtual entities running, both kinds of virtualization alternatives behave similarly in idle state and in CPU/Memory stress test. Contrarily, the results on network performance show differences between the two technologies.

## I. Introduction

Recent advances in virtualization technologies are driving a growing adoption of server virtualization to increase the capacity of data centers. The number of solutions, which enable server consolidation, application isolation, and hardware resources optimization, is increasing fast.

As an example, the use of containers has remarkably increased recently mainly due to the adoption of technologies like Docker [19], which have revolutionized the concept of server virtualization that is usually strictly associated with the concept of hypervisors. There are already several cloud services providers, which make use of container solutions to offer their services like in [9-12].

Closely associated with the development of larger data centers, there is the contribution of those in the growth of energy consumption in the Information and Communication Technology (ICT) sector. In a recent report issued by the National Renewable Energy Laboratory (NREL), it is estimated that U.S. data center usage was approximately 91 million MWh, only in 2013 [8]. Consequently, the study of virtualization technologies power metering represents a key aspect to reducing power consumption and suggests operational optimizations toward an energy-efficient data center design.

The purpose of this paper is to study the capacity of emerging virtualization solutions – such as containers – to reduce the power consumption if compared with hypervisors.

The rest of the paper is organized as follows. In Section II, we summarize background information about the virtualization technologies employed in our analysis. Section III describes in detail the experimental setup and the methodology adopted to carry out our empirical investigation. In Section IV and V, we compare and analyze the achieved results, respectively. Related work is discussed in section VI. Finally, Section VII concludes the paper providing final remarks and future work.

## II. Background

In this section, we provide an overview of the different virtualization environments included in the performance comparison: hypervisors and containers.

Before going through this description, our original idea was to include another emerging virtualization technology such as LXD [13], which is intended to incorporate the most useful aspects of both containers and Virtual Machines (VMs). Unfortunately, because some bugs in the LXD API, it has not been possible to configure the software properly to make a fair comparison with the other technologies.

### A. Hypervisor-based virtualization

Hypervisor-based virtualization operates at the hardware level, thus supporting standalone VMs that are independent and isolated of the host system. As the hypervisor isolates the VM from the underlying host system, e.g. a Linux host machine can run Windows as a guest machine. This is not possible with containers. Also, considering that a full operating system is installed on a virtual machine, the hypervisor-based image will be substantially larger. Emulation of the virtual hardware device incurs more overhead as stated in our previous work [27]. Several open source and commercial hypervisor solutions exist. The ones chosen in our analysis are KVM [14] and Xen [15]. These platforms use two different approaches to perform virtualization by means of hypervisors [16]: *Full virtualization* (KVM), and *Paravirtualization* (Xen). The main difference between the two paradigms is the different level of abstraction that guest systems have of the underlying physical system. With full virtualization, the guest OS – or application – is not aware to be part of a virtualized environment, behaving, as it would be a stand-alone system. Differently, with paravirtualization each VM makes use of hardware that is similar but not identical to the underlying physical hardware, keeping a certain level of dependency from the host system.

This work is partially funded by the FP7 Marie Curie Initial Training Network (ITN) METRICS project (grant agreement No. 607728). The author would like to thank the Cloud team of NomadicLab for providing help and support. Accepted at 8[th] IEEE/ACM International Conference on Utility and Cloud Computing (IEEE/ACM UCC 2015). ©2015 IEEE

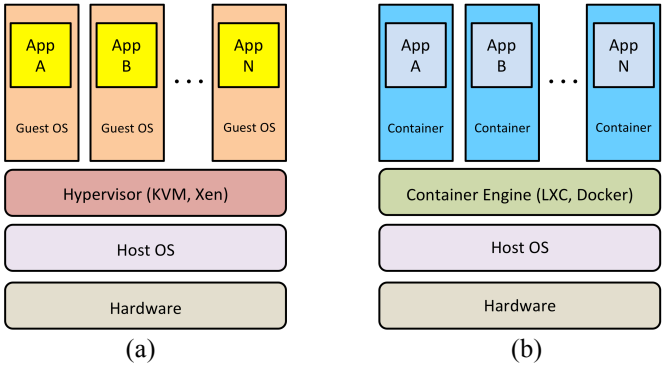

Fig. 1. Virtualization Architecture: (a) hypervisor-based; (b) container-based.

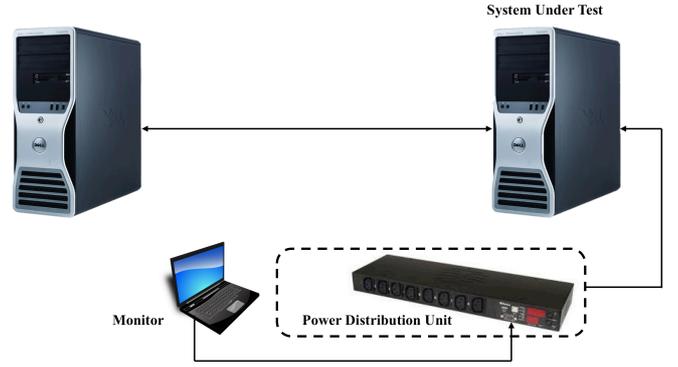

Fig. 2. Testbed setup.

Another difference between KVM and Xen is related to the presence of a privileged domain "Domain 0" in Xen. This domain – which is started by the Xen hypervisor during the boot phase – manages the remaining and unprivileged domains. It works like a "console" that, with special privileges, runs the Xen management tool with direct access to the host hardware. The presence or absence of the "Host OS" in the hypervisor-based virtualization stack (Fig.1a), introduces another classification within this technology. Indeed, the native or bare-metal hypervisors (*type-1 hypervisors*), operate on top of the host's hardware; the hosted hypervisors (*type-2 hypervisors*), operate on top of the host's operating system [17].

### B. Container-based virtualization

Container-based virtualization can be considered as a lightweight alternative to hypervisor-based virtualization (Fig.1b). Containers implement isolation of processes at the operating system level of the underlying host machine, thus avoiding the overhead due to virtualized hardware, and virtual device drivers. One or more processes can be run inside each container. This introduces two advantages: a higher density of virtualized instances, and a smaller disk image. The concept of *"containerization"* is not new in virtualization, but it has achieved much more relevance and real-world adoption recently with the advent of Docker. In this work, we compare Docker performance with LXC [18], which is another alternative in the container-solution family. During the first releases Docker made use of LXC as execution driver, but starting with the version 0.9, *Docker.io* has dropped LXC as default execution environment, replacing it with their own *libcontainer* [20]. Therefore, it is interesting to understand if such change affects the overall performance.

### III. EXPERIMENTAL SETUP AND METHODOLOGY

This section introduces the methodology and a detailed description of the testbed environment illustrated in Figure 2.

**System Under Test (SUT).** The system includes two identical server machines with the following characteristics: computer model *Dell Precision T5500*, with *Intel Xeon X5560* processor (8M Cache, 2.80 GHz, 4 physical cores, 8 threads), 12 GB memory (3x4GB – 1333 MHz DDR3), and a 10 Gbps Network Interface Card.

The operating system used in the host machines is Ubuntu 14.04 LTS (64 bit – kernel version 3.13.0-32-generic). The two machines are directly connected without any switch in between. The choice of performing the network test by using NIC of 10 Gbps is due to the fact that today's application requires more bandwidth, and the total requests coming to servers in data centers is growing drastically [21]. This implies that 1Gbps NIC might be unable to satisfy all the requirements, thus a 10 Gbps Ethernet provides higher bandwidth, which is closer to production scenarios.

**Power Measurement Device.** Power is measured with an external *Power Distribution Unit (PDU)* Raritan [22], with accuracy ±1%. Such devices are widely used in real data centers. In particular, the unit used in our testbed allows remotely controlling several server machines and, at the same time, monitoring power consumption of each machine connected to the PDU. The power information is provided through network cable in real time to an external *Monitor* configured in another host. The PDU measures the *Active Power*, which can be defined as the mean value of the instantaneous power $p$, over one time period T.

$$P = \frac{1}{T} \int_0^T p(t)\, dt$$

In our case, T is 15 seconds, which represents the lowest temporal granularity of the power measurements tool at our disposal. According to [6] and [7], the sample rate $t$ must vary between 1 and 5 seconds. In particular, Chen et al. [7] indicate that the sampling interval should be 5 seconds for stable applications, and 1 second for dynamic workload applications. The sample rate $t$ of our PDU is 3 seconds, which represents a good compromise between the recommended requirements.

**Setup for Virtualized Environment.** The setup used to customize the virtualized environment is similar between all the technologies under evaluation. We made this choice in order to ensure a fair analysis. We perform all the measurements using up to eight guest domains (Dom1 to Dom8). Each virtual guest allocates 2 vCPUs and 1 Gb of memory. The operating system in the guest machines is Ubuntu 14.04 LTS (64 bit). As Hypervisors, we use Xen version 4.4.2, while KVM uses the version 2.0.0 of QEMU. As container, we use the version 1.5.0 of Docker, and the version 1.0.6 of LXC.

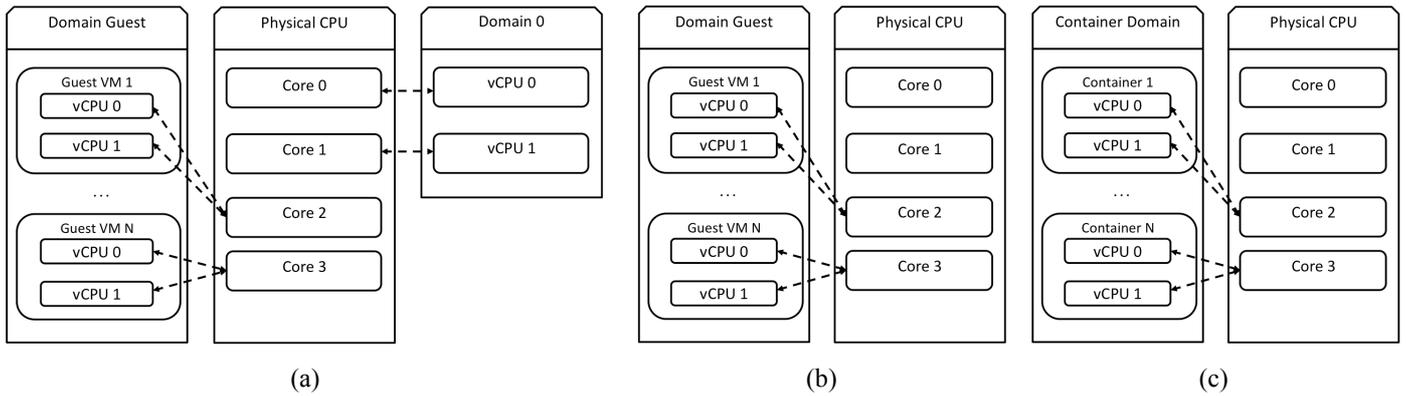

Fig. 3. CPU affinity setup. (a) Xen setup. (b) KVM setup. (c) Container (LXC/Docker) setup.

**vCPU pinning.** In virtualization, v*CPU pinning* (or *Processor affinity*) indicates the possibility to dedicate a physical CPU to a particular virtual CPU or a set of virtual CPUs. It represents a relevant factor, which may affect the results of power evaluation. Several possible configurations exist that can be implemented. The one chosen for our analysis is depicted in Figure 3. In Xen, Domain-0 can run on two CPU cores, while all the other virtual machines (or domains) are pinned with the remaining ones (Fig. 3a). In KVM, we configure the mapping in a differently way, due to the lack of the Domain-0 (Fig. 3b). We make a similar configuration in the container setup (Fig. 3c).

**Network configuration.** Figure 4 shows the network configuration used for our tests. We decided to use the same kind of setup for each virtualized environment. The NIC of all the running virtual entities (Virtual Machine or Container) share the same network bridge, which in turn is mapped to the physical Ethernet of the physical host. Obviously, each technology under evaluation performs network operations (e.g., packet forwarding, packet buffering, scheduling, etc.) in a way that is dependent on the different design and implementation of the virtualization engine (hypervisor or container), and this may generate a different impact on the performance.

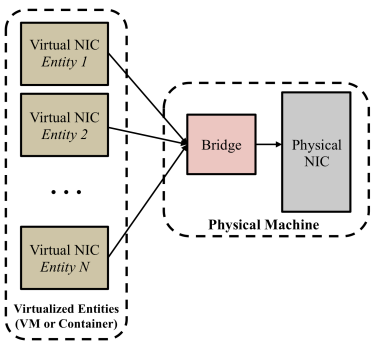

Fig. 4. Network configuration with a shared bridge

**Workloads.** We use benchmark tools with different characteristics to generate different types of workloads. We make this choice in order to challenge a specific hardware segment in our SUT. CPU, memory and, NIC are the main hardware components that we want to test in our experiments. We employ the following tools:

- idle. We use the Unix command line sleep to generate idle state in the system. More specifically, sleep is a package that suspends program execution for a specified period of time.

- sysbench [23]. This benchmark tool allows performing multi-threaded tests for evaluating different parameters (file I/O performance, memory allocation, transfer speed, etc.) under intensive load.

- iperf [24]. This benchmark tool has predefined tests to measure network performance between two hosts. It allows performing bidirectional data transfer, and generating of TCP and UDP traffic.

IV. POWER MEASUREMENTS RESULTS

This section presents the results of our analysis. As explained previously, the selected benchmarks measure *Idle State*, *CPU*, *Memory*, and *Network I/O* performance. Results are organized in four different subsections.

With the aim of improving the readability of the paper, we report the most significant results achieved during our analysis. Each measurement test has been repeated at least 20 times and the results show the average value of these measurements. Some of the graphs also show the standard deviation.

*A. Idle State*

Table 1 shows the power consumption of the four different technologies under evaluation in *Idle* state, when eight virtual guests are running simultaneously. As explained in the previous section, this result has been measured by calling *sleep* in all the virtual guests running on top of the SUT. The results are compared with the power consumption of the SUT without any virtualization technology running. It can be observed how all the different virtualization technologies achieve roughly the same average power consumption in the range of 4 watts, which can be considered – on small scale – almost negligible. Xen consumes most power, even though the difference respect to containers is not high.

TABLE I.

| Platform | Active Power Consumption |
|---|---|
| Native | 123 Watts |
| Xen | 128 Watts |
| KVM | 126 Watts |
| Docker | 124 Watts |
| LXC | 124 Watts |

## B. CPU performance

We test CPU performance with *sysbench*. This particular stress test is designed in such a way to challenge the CPU by calculating prime numbers. The computation is made dividing the number with sequentially increasing numbers. Then, it is verified that the remainder (modulo calculation) is zero. For this particular case, we observe how the power consumption increases when the number of virtual entities allocated in different physical core increases – with respect to the CPU pinning as explained earlier. This is the reason why the graph shows the Active Power when one, two, four, and eight virtual entities are running simultaneously. Figure 5 shows that none of the four virtualization technologies outperform each other. When a single physical core is active (1 virtual guest case), all the platforms consume around 158 Watts. As soon as another virtual entity is allocated to another physical core, we observe an increase of 10 Watts. With four and eight virtual guests equally allocated on two physical cores, the average consumption for hypervisors and containers is around 185 Watts and 190 Watts, respectively.

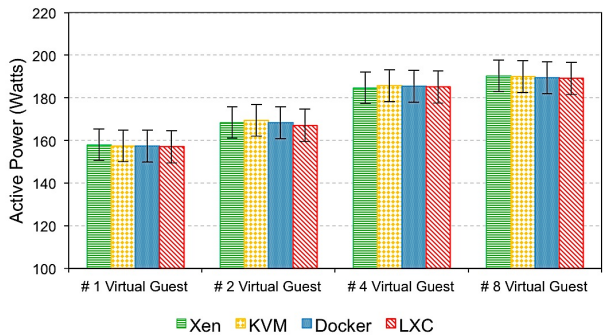

Fig. 5. CPU power consumption with eight active VMs/containers.

## C. Memory Performance

In addition to testing CPU performance, we also used *sysbench* in order to test power consumption of memory. For each technology under evaluation, the results are shown in Figure 6. Even for this particular case, the graph shows the output when one, two, four, and eight VMs or containers are running simultaneously. Both hypervisors and containers behave – on average – similarly and no clear difference can be noticed between the platforms. Considering a balanced distribution of virtual guests between two physical cores, an increase of power consumption from 212 Watts (two virtual guests) to 222 Watts (eight virtual guests) can be noticed.

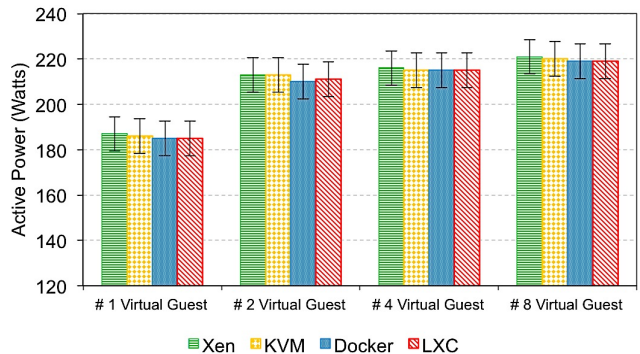

Fig. 6. Memory power consumption with eight active VMs/containers

## D. Network Performance

We used the network benchmarking tool iperf for all of the network performance tests, running at least ten simultaneous iperf sessions per VM/container. This is needed because a single session cannot saturate a 10 Gbps link even for low CPU workload [30, 21]. Our network performance analysis covers several case studies, which are explained below. With the exception of the last case study, the entire network test is performed on a single virtual machine/container.

- *Standard MTU and Jumbo frames performance*. Similarly as in another 10 Gbps networking performance analysis [21], the use of Jumbo frames is mainly due to the impossibility of saturate a 10 Gbps link with standard MTU (1500 bytes). Jumbo frames allow achieving higher throughput by reducing processing overhead. The goal of our tests with Jumbo frames is trying to evaluate how efficiently the virtualization technologies, in conditions of equal throughput, handle this different kind of frames. To enable Jumbo frames, we set the MTU to 9000 bytes in all the physical and virtual NIC involved in the communication. Indeed, as a prerequisite of proper functioning, it is important that no MTU mismatch exists along the link.

- *Bidirectional test*. In this test, the SUT runs both the *iperf server* and the *iperf client*. The decision to execute the test in both directions is due to the fact that TCP has different code paths for send and receive [29].

- *Network performance scalability test*. This test evaluates the power consumption while multiple VMs/containers are running simultaneously. As explained in Section III, we configured our testbed in such a way that several virtual machines/containers share the same physical 10 Gbps NIC. The results of this particular case represent a scenario in which all the virtual entities are stressing the network.

Figure 7 shows the power consumption when the SUT acts as the receiver, and *iperf server* is running in each platform. The incoming traffic is TCP. With MTU 1500 bytes, container-based technologies have the lowest power consumption. Docker consumes around 176 watts on the average, while LXC 177 watts. KVM completes the network operations by consuming around 8 watts more compared to containers. We achieve similar results when we use jumbo frames. The results of Xen are not reported since it was not possible to saturate the 10 Gbps link with a standard network configuration. This issue related to Xen is confirmed even in [28]. Contrarily to the previous case, with higher MTU dimension, Xen is now able to saturate the link and the power consumption is on the same level as KVM. No clear differences in respect to the previous case can be observed with container technologies. Consequently, handling standard frame or jumbo frames does not produce any valuable difference in terms of power consumption, in a single VM or a container.

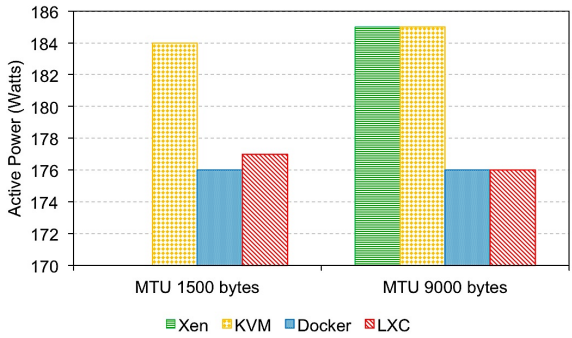

Fig. 7. Power consumption of a single VM/container, which is receiving TCP packet frames with MTU of 1500-bytes and 9000-bytes.

In the tests where the SUT acts as the transmitter, the *iperf client* is running within each virtualized platform. The results from this case are shown in Figure 8. For TCP traffic, all systems are able to transfer an average rate of 9.40 Gbps. KVM, Docker, and LXC have the lowest power consumption. Xen consumes 10 watts more on the average. We achieve similar results when UDP traffic is transmitted, but Xen introduces an even bigger gap than in the TCP case. In fact, the difference to all the other technologies under evaluation varies from 22 to 25 watts.

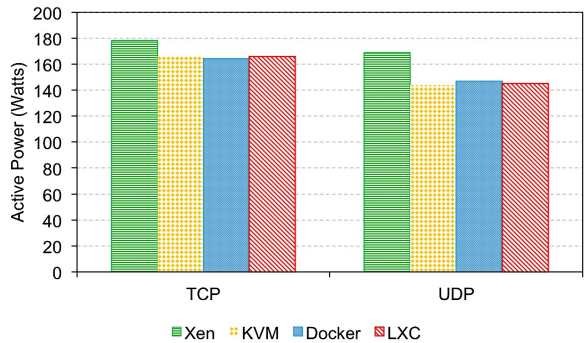

Fig. 8. Power consumption of a single VM/container, which is sending TCP and UDP traffic with MTU of 1500-bytes.

Figure 9 and 10 shows the extreme case of eight virtual entities running simultaneously.

When the virtual entities are receiving TCP traffic using *iperf server*, we observe a difference between the power consumed by hypervisors, and the power consumed by containers (Fig. 9). Xen and KVM consume 199 and 197 watts on the average respectively; Docker is the technology that consumes less power, but LXC is nearly close. The measured difference is approximately 15 watts.

If the VMs/containers are acting as clients generating TCP traffic (Fig. 10), we can observe a slight difference in the obtained result when comparing with the above-described case. The hypervisor performance is now different and KVM consumes less power than Xen. Docker and LXC draw the same amount of power in this particular network case.

## V. ANALYSIS

Our empirical results disclose the following insights about the impact of virtualization on server power consumption usage.

*(i) Idle state Performance.* The insight from this result is that both hypervisors and containers employ an optimized use of

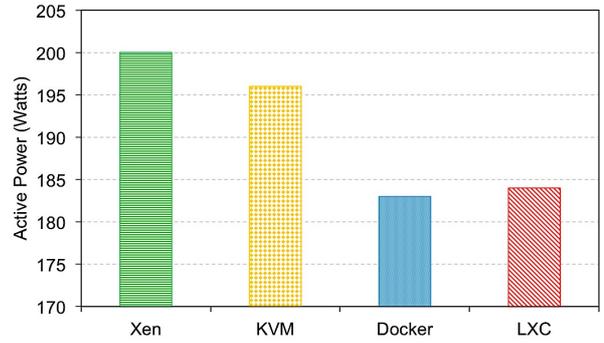

Fig. 9. Power consumption of multiple VMs/containers, which are receiving TCP traffic (MTU 1500-bytes)

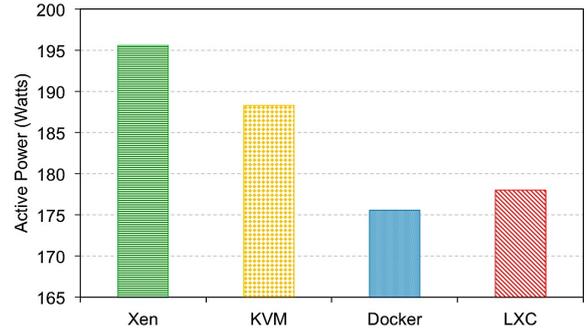

Fig. 10. Power consumption of multiple VMs/containers, which are sending TCP traffic (MTU 1500-bytes)

the Linux power saving system. Moreover, the Domain-0 in Xen does not introduce a significant overhead.

*(ii) CPU and Memory Performance.* The results of the CPU and Memory stress test clearly indicate that no noticeably difference exists in the power consumption of different virtualization designs for CPU heavy workloads, regardless of the number of virtual instances running.

*(iii) Network Performance.* Although CPU and memory performance shows a similar behavior between hypervisors and containers, our network performance analysis shows different results through the different case studies. This is mainly due to the fact that the network packets have to be processed by extra layers in a hypervisor environment in comparison to a container environment. Nevertheless, between the two hypervisors, KVM seems to perform slightly better then Xen. This is probably due to the fact that in Xen when a packet is processed, the physical host NIC will deliver it to the Domain-0 first. After a first processing, it will be transferred to the Domain-N. These additional operations, due to the presence of the Domain-0, can generate additional overhead, which justifies the increase in power consumption.

## VI. RELATED WORK

Earlier works on power modeling of virtualized environment can be categorized in two main areas: analysis based on empirical studies through direct measurements, and analytical power estimation models. Considering the nature of our work, this section provides information only about investigations similar to our work.

Xu et al. in [1] provide a performance comparison between KVM and Xen. The authors conducted several experiments to examine the energy consumption of the two different platforms considering different network traffic patterns and CPU affinity configurations. A similar empirical study that includes OpenVZ among the technologies under evaluation can be found in [2]. The authors discover that an adaptive packet buffering in KVM can reduce the energy consumption caused by network transaction. Jin et al. [3] evaluate the impact of server virtualization in terms of energy efficiency by using several configurations and two different hypervisors. They observe that the energy overhead depends on the type of used hypervisor, and the particular configuration chosen. *Joulemeter* is a solution introduced by Kansal et al. [4]. Without using auxiliary hardware equipment – or any software integration – the authors propose different *"power models to infer power consumption from resource usage at runtime and identify the challenges that arise when applying such models for VM power metering"*. Finally, a recent paper proposes a real-time power estimation of software processes running on any level of virtualization [5] by using an application-agnostic power model.

Expect for the work of Shea et al. [2], none of the related work include Container-based platforms in their analysis and, even more importantly, Docker. Compared with [2], our work includes a wider range of parameters in the network performance analysis.

## VII. CONCLUSION AND FUTURE WORK

In this paper, we have described an empirical investigation in order to evaluate the impact of different virtualization technologies for the power consumption of servers. By performing several measurements experiments and adopting different workloads, we have been able to achieve a better understanding on how much power consumption overhead is introduced by different virtualization solutions: hypervisors (i.e., Xen and KVM), and containers (Docker, LXC). The results show that the power consumption of hypervisors and containers is similar when challenged with heavy CPU and memory workloads. Some differences can be observed in the case of network performance analysis, where, for most of the cases, both of the container solutions introduce lower power consumption.

As further work, it would be interesting to replicate the same kind of measurements with other emerging virtualization technologies such as LXD, once these technologies reach stability so as to ensure a fair comparison. Another relevant aspect, which can be further investigated, is the analysis of power consumption behavior during live migration of VMs/containers. Obviously, our analysis made in this work can be extended to compare different CPU pinning and network configurations.